\documentclass[10pt,twocolumn,english,journal]{IEEEtran}
\usepackage[T1]{fontenc}
\usepackage[latin9]{inputenc}
\usepackage[letterpaper]{geometry}
\usepackage{float}
\usepackage{amsmath}
\usepackage{amssymb}
\usepackage{stackrel}
\usepackage{graphicx}
\usepackage{esint}

\makeatletter

\floatstyle{ruled}
\newfloat{algorithm}{tbp}{loa}
\providecommand{\algorithmname}{Algorithm}
\floatname{algorithm}{\protect\algorithmname}

\usepackage{cite}\usepackage{algorithm}\usepackage{algorithmicx}\usepackage{algpseudocode}\usepackage{babel}

\allowdisplaybreaks \allowdisplaybreaks[4]
\usepackage{flushend}

\usepackage{babel}

\makeatother

\usepackage{babel}
\begin{document}

\title{Approximate Message Passing with Nearest Neighbor  Sparsity Pattern
Learning }

\author{Xiangming Meng, Sheng Wu, Linling Kuang, Defeng (David) Huang, and
Jianhua Lu, \IEEEmembership{Fellow, IEEE} %
\thanks{This work was partially supported by the National Nature Science Foundation
of China (Grant Nos. 91338101, 91438206, and 61231011), the National
Basic Research Program of China (Grant No. 2013CB329001).

X. Meng and J. Lu are with the Department of Electronic Engineering,
Tsinghua University, Beijing, China. (e-mail: %
\mbox{%
mengxm11%
}@mails.tsinghua.edu.cn; %
\mbox{%
lhh-dee%
}@mail.tsinghua.edu.cn).

S. Wu and L. Kuang are with the Tsinghua Space Center, Tsinghua University,
Beijing, China. (e-mail: %
\mbox{%
thuraya%
}@mail.tsinghua.edu.cn; %
\mbox{%
kll%
}@mail.tsinghua.edu.cn).

Defeng (David) Huang is with the School of Electrical, Electronic
and Computer Engineering, The University of Western Australia, Australia
(e- mail: huangdf@ee.uwa.edu.au).%
}}
\maketitle
\begin{abstract}
We consider the problem of recovering clustered sparse signals with
no prior knowledge of the sparsity pattern. Beyond simple sparsity,
signals of interest often exhibits an underlying sparsity pattern
which, if leveraged, can improve the reconstruction performance. However,
the sparsity pattern is usually \textit{unknown} a priori. Inspired
by the idea of k-nearest neighbor (k-NN) algorithm, we propose an
efficient algorithm termed approximate message passing with nearest
neighbor sparsity pattern learning (AMP-NNSPL), which learns the sparsity
pattern adaptively. AMP-NNSPL specifies a flexible spike and slab
prior on the unknown signal and, after each AMP iteration, sets the
sparse ratios as the average of the nearest neighbor estimates via
expectation maximization (EM). Experimental results on both synthetic
and real data demonstrate the superiority of our proposed algorithm
both in terms of reconstruction performance and computational complexity.\end{abstract}


\section{Introduction}

Compressed sensing (CS) aims to accurately reconstruct sparse signals
from undersampled linear measurements\cite{Donoho-CompressiveSensing,Candes-Introduction-to-CS,eldar2012compressed}.
To this end, a plethora of methods have been studied in the past years.
Among others, approximate message passing (AMP) \cite{donoho2009message}
proposed by Donoho \textit{et al.} is one state-of-the-art algorithm
to address sparse signal reconstruction in CS. Moreover, AMP has been
extended to Bayesian AMP (B-AMP) \cite{donoho2010message,krzakala2012probabilistic}
and general linear mixing problems\cite{rangan2011generalized,schniter2011message,Rangan-AMP-Learning}.
While many practical signals can be described as sparse, they often
exhibit an underlying structure, e.g., the nonzero coefficients occur
in clusters\cite{yuan2006model,cevher2009recovery,stojnic2009reconstruction,baraniuk2010model,huang2010benefit,eldar2009robust,eldar2010block}.
Exploiting such intrinsic structure beyond simple sparsity can significantly
boost the reconstruction performance\cite{huang2010benefit,eldar2009robust,eldar2010block}.
To this end, various algorithms have been proposed, e.g., group LASSO\cite{yuan2006model},
StructOMP\cite{huang2011learning}, Graph-CoSaMP\cite{hegde2015nearly},
and block sparse Bayesian learning (B-SBL)\cite{wipf2007empirical,zhang2011sparse,zhang2013extension},
etc. However, these algorithms require knowledge of sparsity pattern
which is usually \textit{unknown} a priori. To reconstruct sparse
signals with unknown structure, a number of methods\cite{he2009exploiting,som2012compressive,yu2012bayesian,andersen2015spatio,fang2015pattern,fang2015two,Yu2015model}
have been developed to use various structured priors to encourage
both sparsity and cluster patterns simultaneously. The main effort
of these algorithms lies in constructing a hierarchical prior model,
e.g., Markov tree\cite{som2012compressive}, structured spike and
slab\cite{yu2012bayesian,andersen2015spatio}, hierarchical Gamma-Gaussian\cite{fang2015pattern,fang2015two,Yu2015model}
to encode the structured sparsity pattern. 

In this letter, we take an alternative approach and propose an efficient
message passing algorithm, termed AMP with nearest neighbor sparsity
pattern learning (AMP-NNSPL), to recover clustered sparse signals
adaptively, i.e., without any prior knowledge of the sparsity pattern.
For clustered sparse signals, if the nearest neighbors of one element
are zeros (nonzeros), it will tend to be zero (nonzero) with high
probability, a similar idea of k-nearest neighbor (k-NN) algorithm
which assumes that data close together more likely belong to the same
category\cite{fix1951discriminatory,cover1967nearest}. Therefore,
instead of explicitly modeling the sophisticated sparsity pattern,
AMP-NNSPL specifies a flexible spike and slab prior on the unknown
signal and, after each AMP iteration, updates the sparse ratios as
the average of their nearest neighbor estimates via expectation maximization
(EM)\cite{dempster1977maximum}. In this way, the sparsity pattern
is learned adaptively. Simulations results on both synthetic and real
data demonstrate the superiority of our proposed algorithm both in
terms of reconstruction performance and computational efficiency.

\section{\label{sec:System-Model}System Model }

Consider the following linear Gaussian model
\begin{equation}
\mathbf{y}=\mathbf{Ax}+\mathbf{w},\label{eq:linear_model}
\end{equation}
where $\mathbf{x}\in\mathbb{R}^{N}$ is the unknown signal, $\mathbf{y}\in\mathbb{R}^{M}$
is the available measurements, $\mathbf{A}\in\mathbb{R}^{M\times N}$
is the known measurement matrix, and $\mathbf{w}\in\mathbb{R}^{M}\sim\mathcal{N}\bigl(\mathbf{w};0,\Delta_{0}\mathbf{I}\bigr)$
is the additive noise. $\mathcal{N}\bigl(\mathbf{x};\mathbf{m},\mathbf{C}\bigr)$
denotes a Gaussian distribution of $\mathbf{x}$ with mean $\mathbf{m}$
and covariance $\mathbf{C}$ and $\mathbf{I}$ denotes the identity
matrix. Our goal is to estimate $\mathbf{x}$ from $\mathbf{y}$ when
$M\ll N$ and $\mathbf{x}$ is clustered sparse while its specific
sparsity pattern is unknown a priori. 

To enforce sparsity, from a Bayesian perspective, the signals are
assumed to follow sparsity-promoting prior distributions, e.g., Laplace
prior\cite{park2008bayesian}, automatic relevance determination \cite{tipping2001sparse},
and spike and slab prior\cite{vila2013expectation,krzakala2012probabilistic}.
In this letter we consider a flexible spike and slab prior of the
form
\begin{equation}
p_{0}(\mathbf{x})=\stackrel[i=1]{N}{\prod}p_{0}(x_{i})=\stackrel[i=1]{N}{\prod}\bigl[\bigl(1-\lambda_{i}\bigr)\delta(x_{i})+\lambda_{i}f(x_{i})\bigr],\label{eq:common_BG_prior}
\end{equation}
where $\lambda_{i}\in(0,1)$ is the sparse ratio, i.e., the probability
of $x_{i}$ being nonzero, $\delta(x_{i})$ is the Dirac delta function,
$f(x_{i})$ is the distribution of the nonzero entries in $\mathbf{x}$,
e.g., $f(x_{i})=\mathcal{N}(x_{i};\mu_{0},\tau_{0})$ for sparse Gaussian
signals and $f(x_{i})=\delta(x_{i}-1)$ for sparse binary signals,
etc.

It is important to note that in (\ref{eq:common_BG_prior}) we specify
an individual $\lambda_{i}$ for each entry, as opposed to a common
value in \cite{vila2013expectation,krzakala2012probabilistic}. This
is a key feature that will be exploited by the proposed algorithm
for reconstruction of structured sparse signals. Up to now, it seems
that no structure is ever introduced to enforce the underlying sparsity
pattern. Indeed, if the sparse ratios $\lambda_{i},i=1,\ldots,N$
are learned independently, we will not benefit from the potential
structure. The main contribution of this letter is a novel adaptive
learning method which encourages clustered sparsity, as descried in
Section \ref{sec:Adaptive-Sparse-Reconstruction}.

\section{\label{sec:Adaptive-Sparse-Reconstruction}Proposed Algorithm}

In this section, inspired by the idea of k-NN, we propose an adaptive
reconstruction algorithm to recover clustered sparse signals without
any prior knowledge of the sparsity pattern, e.g., structure and sparse
ratio. 

Before proceeding, we first give a brief description of AMP. Generally,
AMP decouples the vector estimation problem (\ref{eq:linear_model})
into $N$ scalar problems in the asymptotic regime\cite{montanari2012graphical,bayati2011dynamics}

\begin{equation}
\mathbf{y}=\mathbf{Ax}+\mathbf{w}\longrightarrow\begin{cases}
R_{1}=x_{1}+\tilde{w}_{1}\\
\vdots & ,\\
R_{N}=x_{N}+\tilde{w}_{N}
\end{cases}\label{eq:scalar_system_AMP}
\end{equation}
where the effective noise $\tilde{w}_{i}$ asymptotically follows
$\mathcal{N}\bigl(\tilde{w}_{i};0,\Sigma_{i}\bigr)$. The values of
$R_{i},\Sigma_{i}$ are updated iteratively in each AMP iteration
(see Algorithm \ref{AMP-NNSPL Algorithm}) and the posterior distribution
of $x_{i}$ is estimated as
\begin{equation}
q\bigl(x_{i}|R_{i},\Sigma_{i}\bigr)=\frac{1}{Z(R_{i},\Sigma_{i})}p_{0}\bigl(x_{i}\bigr)\mathcal{N}\bigl(x_{i};R_{i},\Sigma_{i}\bigr),\label{eq:post_dist_of_xi}
\end{equation}
where $Z(R_{i},\Sigma_{i})$ is the normalization constant. From (\ref{eq:post_dist_of_xi}),
the estimates of the mean and variance of $x_{i}$ are
\begin{align}
g_{a}(R_{i},\Sigma_{i}) & =\int x_{i}q\bigl(x_{i}|R_{i},\Sigma_{i}\bigr)dx_{i},\label{eq:mean_def}\\
g_{c}(R_{i},\Sigma_{i}) & =\int x_{i}^{2}q\bigl(x_{i}|R_{i},\Sigma_{i}\bigr)dx_{i}-g_{a}^{2}(R_{i},\Sigma_{i}).\label{eq:var_def}
\end{align}

For more details of AMP and its extensions, the readers are referred
to \cite{donoho2009message,donoho2010message,montanari2012graphical,krzakala2012probabilistic}.
Two problems arise in traditional AMP. First, it assumes full knowledge
of the prior distribution and noise variance, which is an impractical
assumption. Second, it does not account for the potential structure
of sparsity. In the sequel, we resort to expectation maximization
(EM) to learn the unknown hyperparameters. Further, to encourage structured
sparsity, we develop a nearest neighbor sparsity pattern learning
rule motivated by the idea of k-NN algorithm. For lack of space, we
only consider the sparse Gaussian case, $f\bigl(x_{i}\bigr)=\mathcal{N}\bigl(x_{i};\mu_{0},\tau_{0}\bigr),$
while generalization to other settings is possible.

The hidden variables are chosen as the unknown signal vector $\mathbf{x}$
and the hyperparameters are denoted by $\boldsymbol{\theta}$. The
specific definition of $\boldsymbol{\theta}$ depends on the choice
of distribution $f\bigl(x\bigr)$ in (\ref{eq:common_BG_prior}).
In the Gaussian case, $\boldsymbol{\theta}=\bigl\{\mu_{0},\tau_{0},\Delta_{0},\lambda_{i},i=1,\ldots,N\bigr\}$
while in the binary case, $\boldsymbol{\theta}=\bigl\{\Delta_{0},\lambda_{i},i=1,\ldots,N\bigr\}$.
Denote by $\boldsymbol{\theta}^{t}$ the estimate of hyperparameters
at the $t$th EM iteration, then EM alternates between the following
two steps\cite{dempster1977maximum} 
\begin{align}
Q\bigl(\boldsymbol{\theta},\boldsymbol{\theta}^{t}\bigr) & =\mathsf{E}\Bigl\{\ln p\bigl(\mathbf{\mathbf{x}},\mathbf{y}\bigr)|\mathbf{y};\boldsymbol{\theta}^{t}\Bigr\},\label{eq:E_step_def-1}\\
\boldsymbol{\theta}^{t+1} & =\arg\underset{\boldsymbol{\theta}}{\max}Q\bigl(\boldsymbol{\theta},\boldsymbol{\theta}^{t}\bigr),\label{eq:max_step_def-1}
\end{align}
where $\mathsf{E}\bigl\{\cdot|\mathbf{y};\boldsymbol{\theta}^{t}\bigr\}$
denotes expectation conditioned on observations $\mathbf{y}$ with
parameters $\boldsymbol{\theta}^{t}$, i.e., the expectation is with
respect to the posterior distribution $p\bigl(\mathbf{\mathbf{x}}|\mathbf{y};\boldsymbol{\theta}^{t}\bigr)$.
From (\ref{eq:linear_model}), (\ref{eq:common_BG_prior}), the joint
distribution $p(\mathbf{\mathbf{x}},\mathbf{y})$ in (\ref{eq:E_step_def-1})
is defined as 
\begin{equation}
p(\mathbf{\mathbf{x}},\mathbf{y})=p\bigl(\mathbf{y}|\mathbf{\mathbf{x}}\bigr)\prod_{i}(1-\lambda_{i})\delta(x_{i})+\lambda_{i}f(x_{i}),\label{eq:joint_dist}
\end{equation}
where $p(\mathbf{y}|\mathbf{\mathbf{x}})=\mathcal{N}\bigl(\mathbf{y};\mathbf{Ax},\Delta_{0}\mathbf{I}\bigr)$.
AMP offers an efficient approximation of $p\bigl(\mathbf{\mathbf{x}}|\mathbf{y};\boldsymbol{\theta}^{t}\bigr)$,
denoted as $q\bigl(\mathbf{\mathbf{x}}|\mathbf{y};\boldsymbol{\theta}^{t}\bigr)=\prod_{i}q\bigl(x_{i}|R_{i},\Sigma_{i}\bigr)$,
whereby the E step (\ref{eq:E_step_def-1}) can be efficiently calculated.
Since joint optimization of $\mathbf{\boldsymbol{\theta}}$ is difficult,
we adopt the incremental EM update rule proposed in \cite{neal1998view},
i.e., we update one or partial elements at a time while holding the
other parameters fixed. 

After some algebra, the marginal posterior distribution of $x_{i}$
in (\ref{eq:post_dist_of_xi}) can be written as 
\begin{equation}
q\bigl(x_{i}|R_{i},\Sigma_{i}\bigr)=\bigl(1-\pi_{i}\bigr)\delta\bigl(x_{i}\bigr)+\pi_{i}\mathcal{N}\bigl(x_{i};m_{i},V_{i}\bigr),\label{eq:marginal_post_x}
\end{equation}
where 
\begin{align}
V_{i} & =\frac{\tau_{0}\Sigma_{i}}{\Sigma_{i}+\tau_{0}},\label{eq:var1}\\
m_{i} & =\frac{\tau_{0}R_{i}+\Sigma_{i}\mu_{0}}{\Sigma_{i}+\tau_{0}},\label{eq:mean1}\\
\pi_{i} & =\frac{\lambda_{i}}{\lambda_{i}+\bigl(1-\lambda_{i}\bigr)\exp\bigl(-\mathcal{L}\bigr)},\label{eq:pi_calculate_formula}\\
\mathcal{L} & =\frac{1}{2}\ln\frac{\Sigma_{i}}{\Sigma_{i}+\tau_{0}}+\frac{R_{i}^{2}}{2\Sigma_{i}}-\frac{\bigl(R_{i}-\mu_{0}\bigr)^{2}}{2\bigl(\Sigma_{i}+\tau_{0}\bigr)}.\label{eq:log_znz_2_zz}
\end{align}
Note that for notational brevity, we have omitted the iteration index
$t$. The mean and variance defined in (\ref{eq:mean_def}) and (\ref{eq:var_def})
can now be explicitly calculated as 
\begin{align}
g_{a}\bigl(R_{i},\Sigma_{i}\bigr) & =\pi_{i}m_{i},\label{eq:post_mean-1-1}\\
g_{c}\bigl(R_{i},\Sigma_{i}\bigr) & =\pi_{i}\bigl(m_{i}^{2}+V_{i}\bigr)-g_{a}^{2}\bigl(R_{i},\Sigma_{i}\bigr).\label{eq:post_var-1}
\end{align}

To learn the sparse ratios $\lambda_{i},i=1,\ldots,N$, we need to
maximize $Q\bigl(\boldsymbol{\theta},\boldsymbol{\theta}^{t}\bigr)$
with respect to $\lambda_{i}$. After some algebra, we obtain the
standard EM update equation as $\lambda_{i}^{t+1}=\pi_{i}^{t}$, which,
albeit simple, fails to capture the inherent structure in the sparsity
pattern. To address this problem, a novel learning rule is proposed
as follows
\begin{equation}
\lambda_{i}^{t+1}=\frac{1}{\bigl|\mathcal{N}\bigl(i\bigr)\bigr|}\underset{j\in\mathcal{N}(i)}{\sum}\pi_{j}^{t},\label{eq:Novel_update_lamda}
\end{equation}
where $\mathcal{N}\bigl(i\bigr)$ denotes the set of nearest neighbor
indexes of element $x_{i}$ in $\mathbf{x}$ (\ref{eq:linear_model})
and $\bigl|\mathcal{N}\bigl(i\bigr)\bigr|$ denotes the cardinality
of $\mathcal{N}\bigl(i\bigr)$. For one dimensional (1D) data , $\mathcal{N}\bigl(i\bigr)=\bigl\{ i-1,i+1\bigr\}$%
\footnote{For end points of 1D data, the nearest neighbor set has only one element.
For edge points of 2D data, the nearest neighbor set has only two
or three elements.%
} and $\bigl|\mathcal{N}\bigl(i\bigr)\bigr|=2$, while for two dimensional
(2D) data, $\mathcal{N}\bigl(i\bigr)=\bigl\{(q,l-1),(q,l+1),(q-1,l),(q+1,l)\bigr\}$
and $\bigl|\mathcal{N}\bigl(i\bigr)\bigr|=4$, where $(q,l)$ indicates
the coordinates of $x_{i}$ in the 2D space. Generalizations to other
cases can be made.

Note that in (\ref{eq:Novel_update_lamda}), we have chosen the nearest
neighbor of each element, excluding itself, as the neighboring set.
The estimate of one sparse ratio is not determined by its own estimate,
but rather the average of its nearest neighbor estimates. The insight
for this choice is that, for clustered sparse signals, if the nearest
neighbors of one element are zero (nonzero), it will be zero (nonzero)
with high probability, a similar idea to k-NN. If the neighboring
set is chosen as the whole elements, the proposed algorithm reduces
to EM-BG-GAMP\cite{vila2013expectation,krzakala2012probabilistic}.

The leaning of other hyperparameters follows the standard rule of
EM algorithm. Maximizing $Q\bigl(\boldsymbol{\theta},\boldsymbol{\theta}^{t}\bigr)$
with respect to $\Delta_{0}$ and after some algebra, we obtain 
\begin{equation}
\Delta_{0}^{t+1}=\frac{1}{M}\sum_{a}\Bigl[\frac{\bigl(y_{a}-Z_{a}^{t}\bigr)^{2}}{\bigl(1+V_{a}^{t}/\Delta_{0}^{t}\bigr)^{2}}+\frac{\Delta_{0}^{t}V_{a}^{t}}{\Delta_{0}^{t}+V_{a}^{t}}\Bigr],\label{eq:noise_learning_free_energy}
\end{equation}
where $Z_{a}^{t}$ and $V_{a}^{t}$ are obtained within the AMP iteration
and are defined in Algorithm \ref{AMP-NNSPL Algorithm}. Similarly,
maximizing $Q\bigl(\boldsymbol{\theta},\boldsymbol{\theta}^{t}\bigr)$
with respect to $\mu_{0}$ and $\tau_{0}$ results in the update equations
\begin{align}
\mu_{0}^{t+1} & =\frac{\sum_{i}\pi_{i}^{t}m_{i}^{t}}{\sum_{i}\pi_{i}^{t}},\label{eq:mu_0_update}\\
\tau_{0}^{t+1} & =\frac{1}{\sum_{i}\pi_{i}^{t}}\underset{i}{\sum}\pi_{i}^{t}\bigl[\bigl(\mu_{0}^{t}-m_{i}^{t}\bigr)^{2}+V_{i}\bigr].\label{eq:tao_0_update}
\end{align}

Valid initialization of the unknown hyperparameters is essential since
EM algorithm may converge to a local maximum or a saddle point of
the likelihood function\cite{dempster1977maximum}. The sparse ratios
$\lambda_{i}$ and noise variance $\Delta_{0}$ are initialized as
$\lambda_{i}^{1}=0.5$ and $\Delta_{0}^{1}=\bigl\Vert\mathbf{y}\bigr\Vert_{2}^{2}/M\bigl(\textrm{SNR}^{0}+1\bigr)$,
respectively, where $\textrm{SNR}^{0}$ is suggested to be 100 \cite{vila2013expectation}.
For the sparse Gaussian case, active mean $\mu_{0}$ and variance
$\tau_{0}$ are initialized as $\mu_{0}^{1}=0$, and $\tau_{0}^{1}=\bigl(\bigl\Vert\mathbf{y}\bigr\Vert_{2}^{2}-M\Delta_{0}^{1}\bigr)/\lambda_{i}^{1}\bigl\Vert\mathbf{A}\bigr\Vert_{F}^{2}$,
respectively, where$\bigl\Vert\mathbf{\cdot}\bigr\Vert_{2}$, $\bigl\Vert\mathbf{\cdot}\bigr\Vert_{F}$
are the $l_{2}$ norm and Frobenius norm, respectively.

The proposed approximate message passing with nearest neighbor sparsity
pattern learning (AMP-NNSPL) is summarized in Algorithm \ref{AMP-NNSPL Algorithm}.
The complexity of AMP-NNSPL is dominated by matrix-vector multiplications
in the original AMP and thus only scales as $\mathcal{O}(MN)$, i.e.,
the proposed algorithm is computationally efficient.

\begin{algorithm}
\protect\caption{AMP-NNSPL Algorithm }

\textbf{Input}: $\mathbf{y}$ $\mathbf{A}$.

\begin{raggedright}
\textbf{Initialization}: Set $t=1$ and $T_{max},\epsilon_{toc}$.
Initialize $\mu_{0},\tau_{0},\Delta_{0}$ and $\lambda_{i},i=1,\ldots,N$
as in Section \ref{sec:Adaptive-Sparse-Reconstruction}. $\hat{x}_{i}^{1}=\int x_{i}p_{0}(x_{i})dx_{i},\nu_{i}^{1}=\int|x_{i}-\hat{x}_{i}^{1}|^{2}p_{0}(x_{i})dx_{i},i=1,\ldots,N$,
$V_{a}^{0}=1,Z_{a}^{0}=y_{a},a=1,\ldots,M.$
\par\end{raggedright}

1) Factor node update: For $a=1,\ldots,M$
\begin{align*}
V_{a}^{t} & =\sum_{i}|A_{ai}|^{2}\nu_{i}^{t},\\
Z_{a}^{t} & =\sum_{i}A_{ai}\hat{x}_{i}^{t}-\frac{V_{a}^{t}}{\Delta_{0}^{t}+V_{a}^{t-1}}\bigl(y_{a}-Z_{a}^{t-1}\bigr).
\end{align*}

2) Variable node update: For $i=1,\ldots,N$
\begin{alignat*}{1}
\Sigma_{i}^{t} & =\Bigl[\sum_{a}\frac{|A_{ai}|^{2}}{\Delta_{0}^{t}+V_{a}^{t}}\Bigr]^{-1},\\
R_{i}^{t} & =\hat{x}_{i}^{t}+\Sigma_{i}^{t}\sum_{a}\frac{A_{ai}\bigl(y_{a}-Z_{a}^{t}\bigr)}{\Delta_{0}^{t}+V_{a}^{t}},\\
\hat{x}_{i}^{t+1} & =g_{a}\left(R_{i}^{t},\Sigma_{i}^{t}\right),\\
\hat{\nu}_{i}^{t+1} & =g_{c}\left(R_{i}^{t},\Sigma_{i}^{t}\right).
\end{alignat*}

3) Update $\lambda_{i}^{t+1},i=1,\ldots N$, as (\ref{eq:Novel_update_lamda});

4) Update $\mu_{0}^{t+1},\tau_{0}^{t+1},\Delta_{0}^{t+1}$ as (\ref{eq:mu_0_update}),
(\ref{eq:tao_0_update}), and (\ref{eq:noise_learning_free_energy});

5) Set $t\leftarrow t+1$ and proceed to step 1) until $T_{max}$
iterations or $\bigl\Vert\mathbf{\hat{x}}^{t+1}-\mathbf{\hat{x}}^{t}\bigr\Vert_{2}<\epsilon_{toc}\bigl\Vert\hat{\mathbf{x}}^{t}\bigr\Vert_{2}$
. \label{AMP-NNSPL Algorithm}
\end{algorithm}

\section{\label{sec:Numerical-Experiments}Numerical Experiments}

In this section, a series of numerical experiments are performed to
demonstrate the performance of the proposed algorithm under various
settings. Comparisons are made to some state-of-the-art methods which
need no prior information of the sparstiy pattern, e.g., PC-SBL\cite{fang2015pattern}
and its AMP version PCSBL-GAMP\cite{fang2015two}, MBCS-LBP\cite{Yu2015model},
and EM-BG-GAMP\cite{vila2013expectation}. The performance of Basis
Pursuit (BP) \cite{chen1998atomic,candes2005decoding,candes2006robust}
is also evaluated. Throughout the experiments, we set the maximum
number of iterations for AMP-NNSPL, PCSBL-GAMP, and EM-BG-GAMP to
be $T_{max}=200$, and the tolerance value of termination to be $\epsilon_{toc}=10^{-6}.$
For other algorithms, we use the defaut setting. The elements of measurement
matrix $\mathbf{A}\in\mathbb{R}^{M\times N}$ are independently generated
following standard Gaussian distribution and the columns are normalized
to unit norm. The success rate is defined as the ratio of the number
of successful trials to the total number of experiments, where a trial
is successful if the normalized mean square error (NMSE) is less than
-60 dB, where $\mathsf{NMSE}=20\log_{10}(\bigl\Vert\mathbf{\hat{x}}-\mathbf{x}\bigr\Vert_{2}/\bigl\Vert\mathbf{x}\bigr\Vert_{2})$.
The pattern recovery success rate is defined as the ratio of the number
of successful trials to the total number of experiments, where a trial
is successful if the support is exactly recovered. A coefficient whose
magnitude is less than $10^{-4}$ is deemed as a zero coefficient.

\subsection{Synthetic Data}

\begin{figure}
\includegraphics[width=3.85cm]{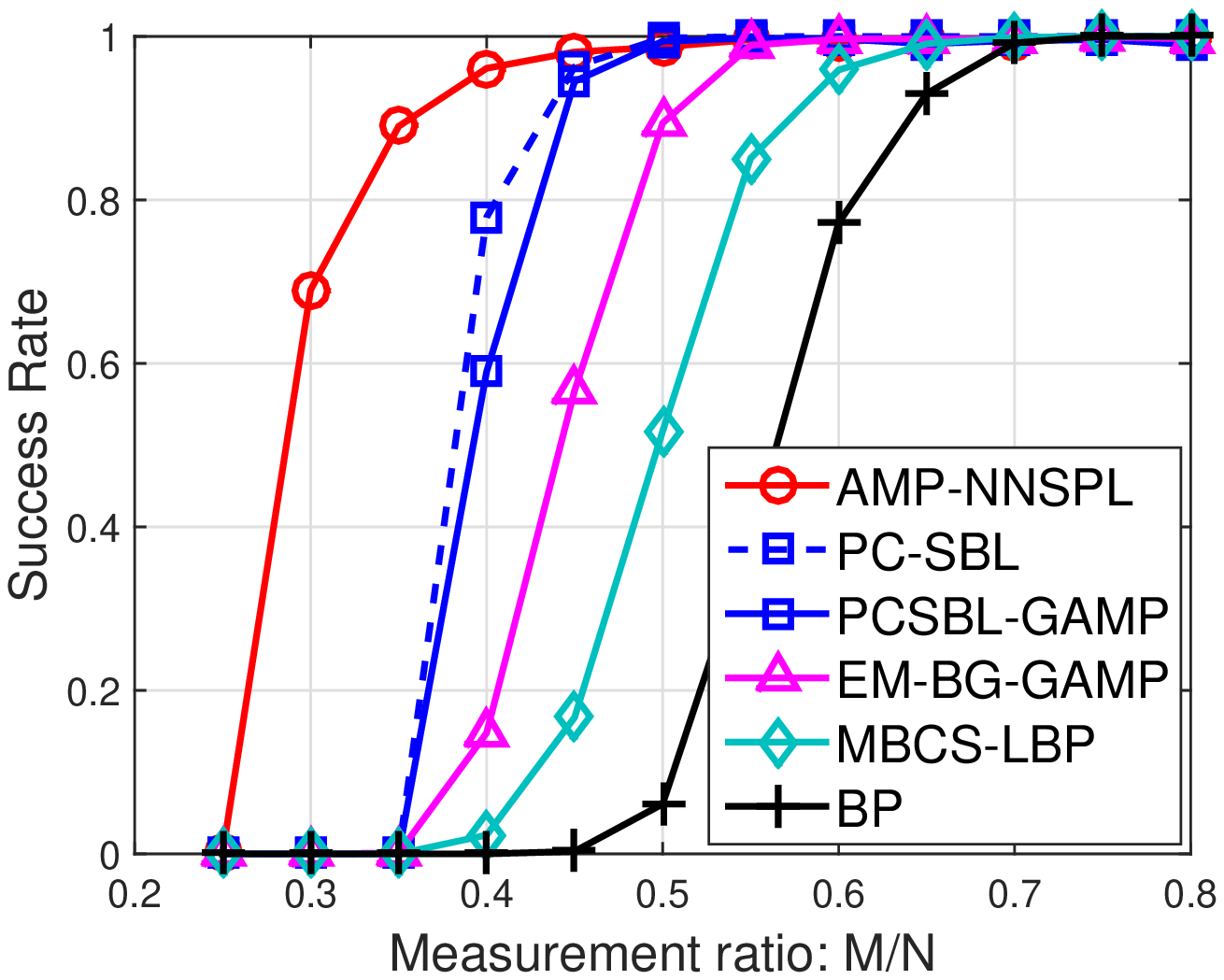}\includegraphics[width=3.85cm]{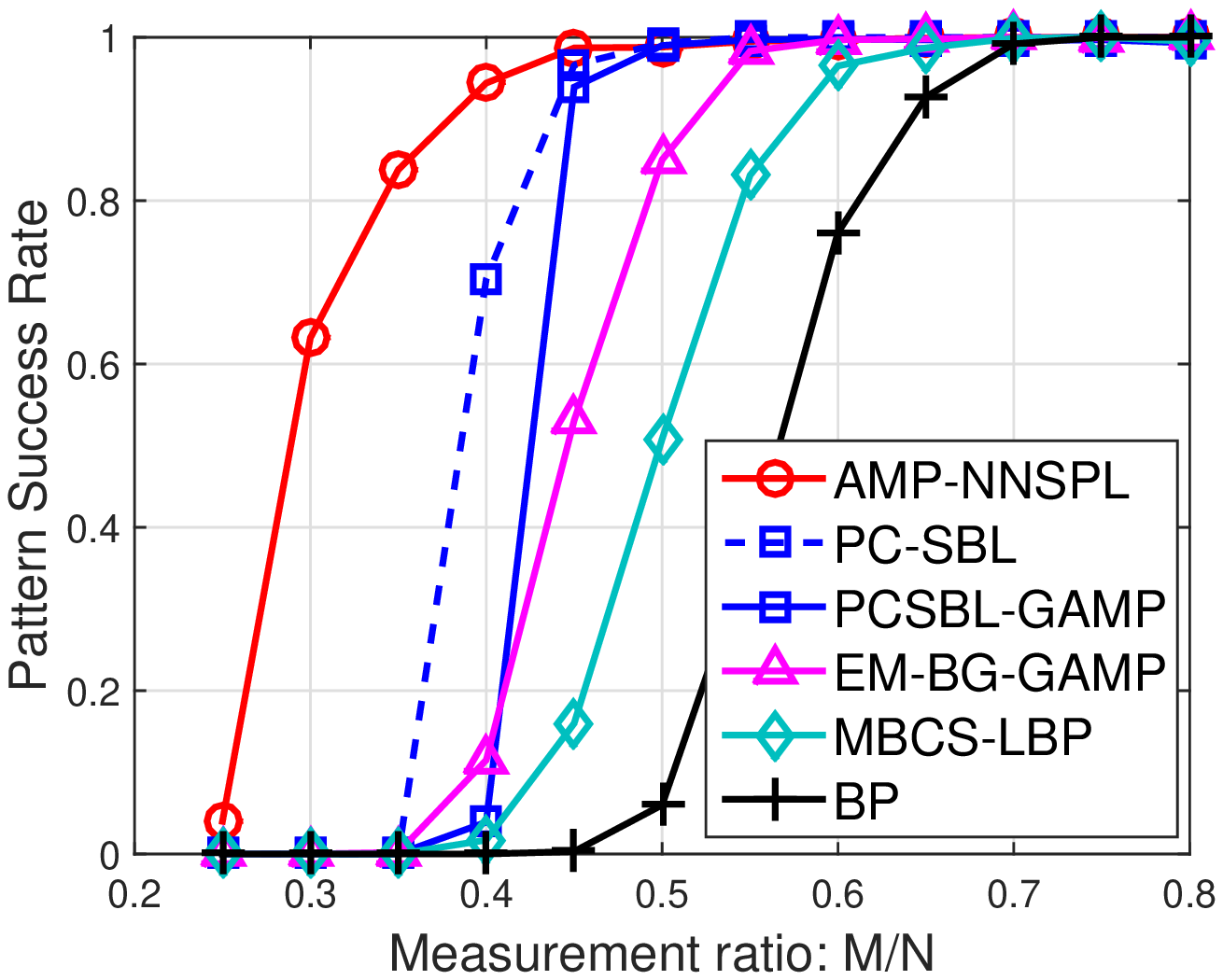}

\protect\caption{Success rate (left) and pattern success rate (right) vs. $M/N$ for
block-sparse signals $N=100$, $K=25$, $L=4$, noiseless case. }

\label{Success_rate_noiseless}
\end{figure}

\begin{figure}
\includegraphics[width=3.85cm]{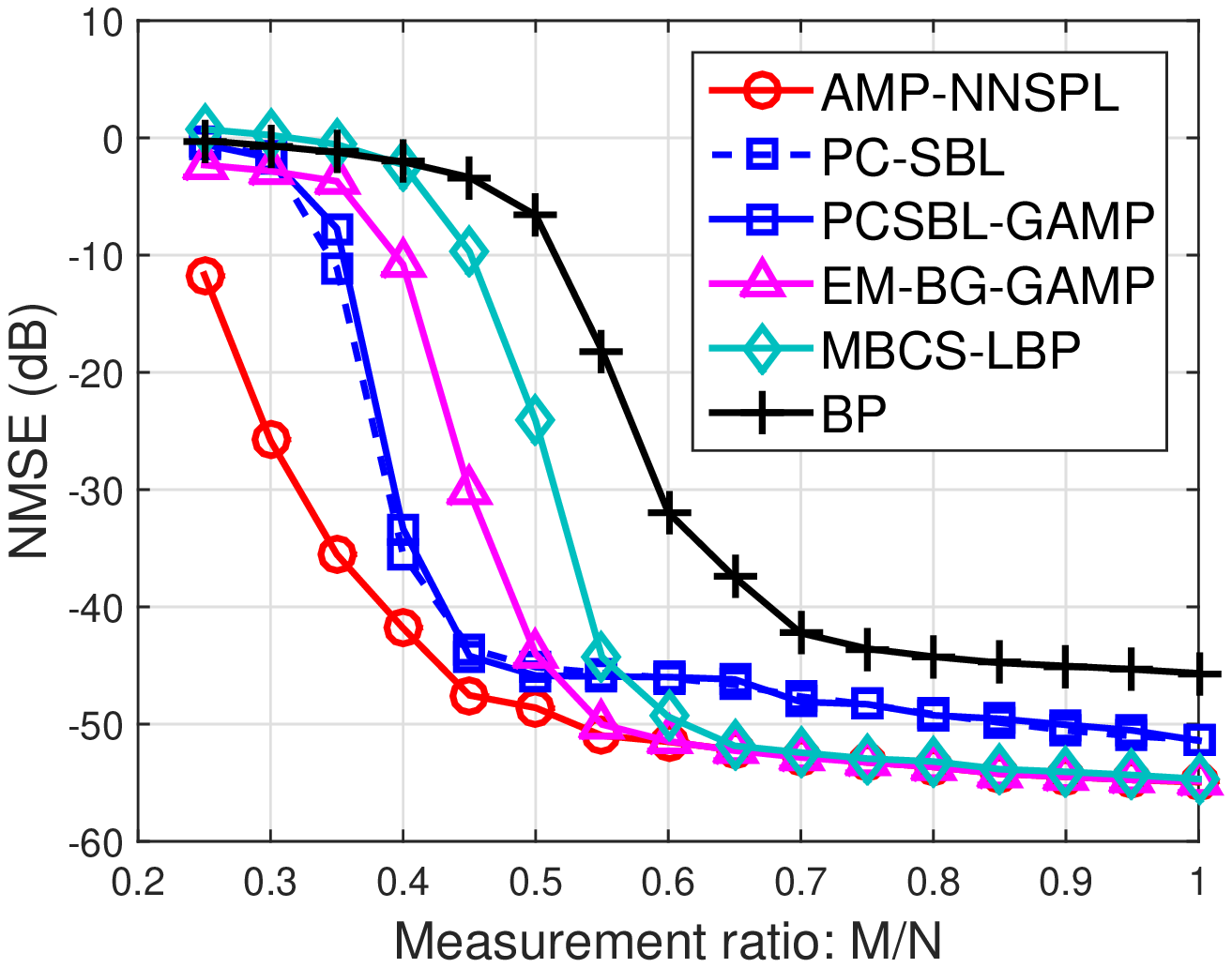}\includegraphics[width=3.85cm]{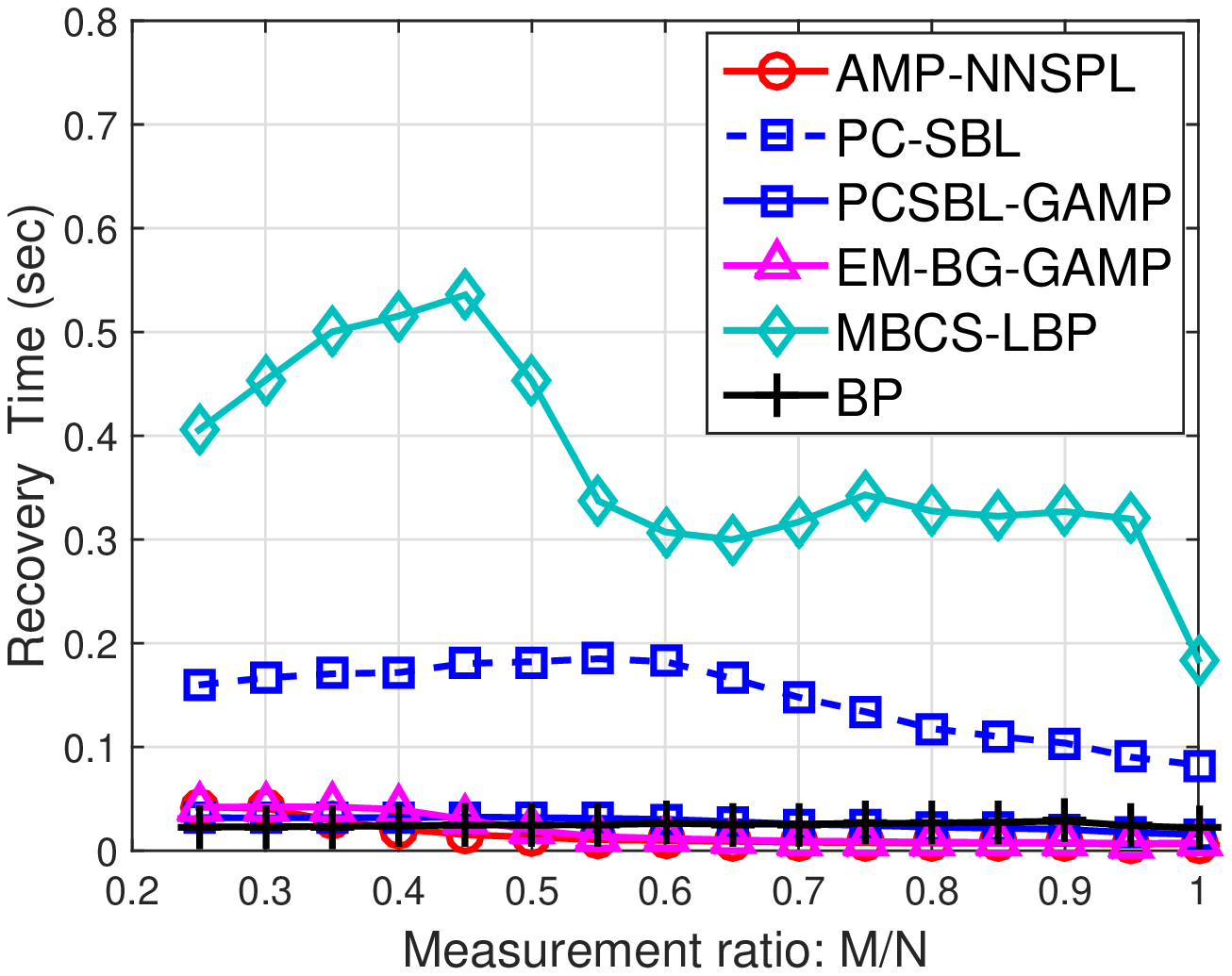}

\protect\caption{NMSE (left) and recovery time (right) vs. $M/N$ for block-sparse
signals $N=100$, $K=25$, $L=4$, SNR = 50 dB.}

\label{Success_rate_noiseless-1}
\end{figure}

We generate synthetic block-sparse signals in a similar way as \cite{zhang2013extension,fang2015pattern},
where $K$ nonzero elements are partitioned into $L$ blocks with
random sizes and random locations. Set $N=100,K=25,L=4$ and the nonzero
elements are generated independently following Gaussian distribution
with mean $\mu_{0}=3$ and variance $\tau_{0}=1$. The results are
averaged over 1000 independent runs. Fig. \ref{Success_rate_noiseless}
depicts the success rate and pattern recovery success rate. It can
be seen that AMP-NNSPL achieves the highest success rate and pattern
recovery rate at various measurement ratios. In the noisy setting,
Fig. \ref{Success_rate_noiseless-1} shows the average NMSE and runtime
of different algorithms when the signal to noise ratio (SNR) is 50
dB, where $SNR=20\log_{10}(\bigl\Vert\mathbf{A}\mathbf{x}\bigr\Vert_{2}/\bigl\Vert\mathbf{w}\bigr\Vert_{2})$.
We see that AMP-NNSPL outperforms other methods both in terms of NMSE
and computational efficiency. 

\subsection{Real Data}

\begin{figure}
\includegraphics[width=4cm]{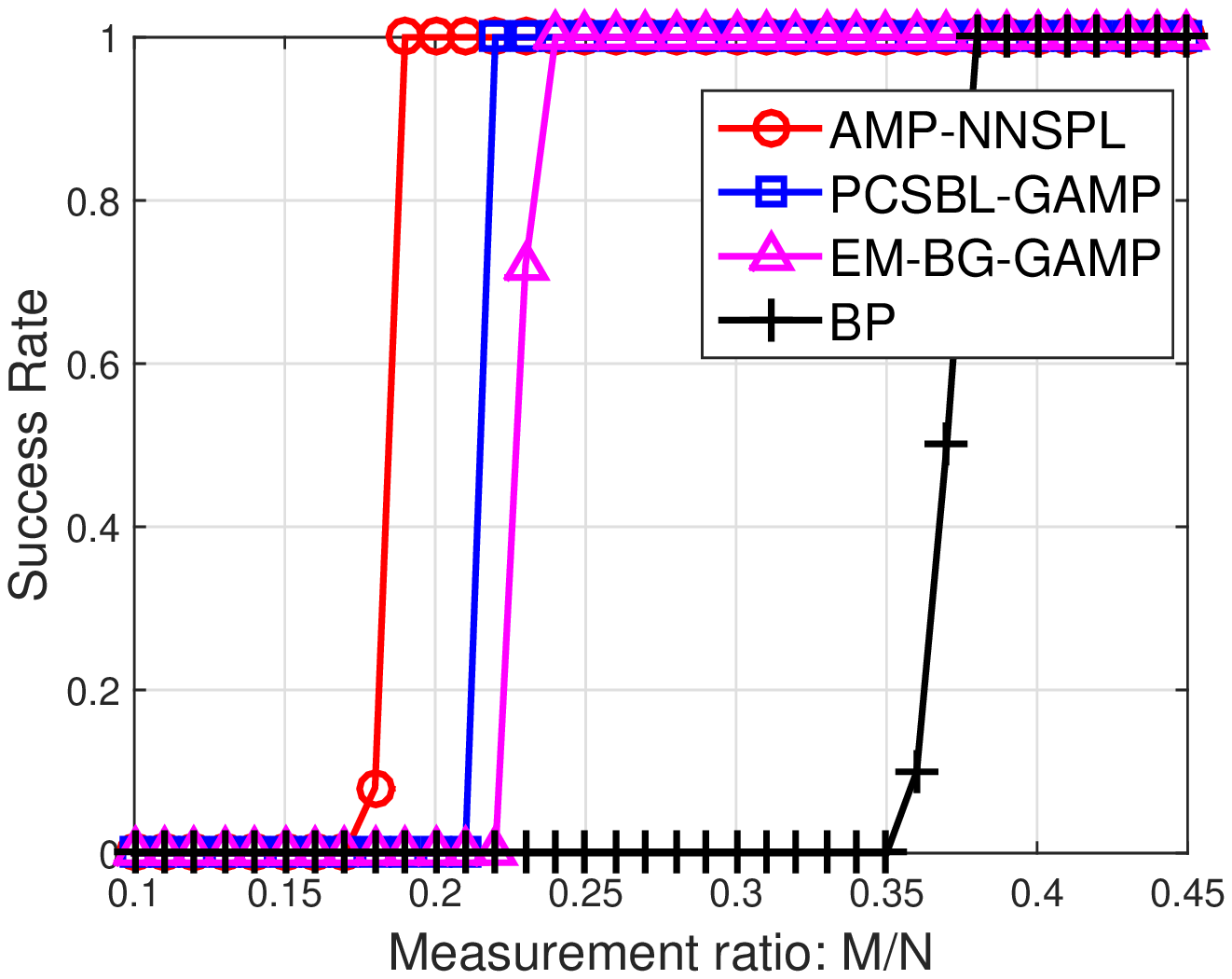}\includegraphics[width=4cm]{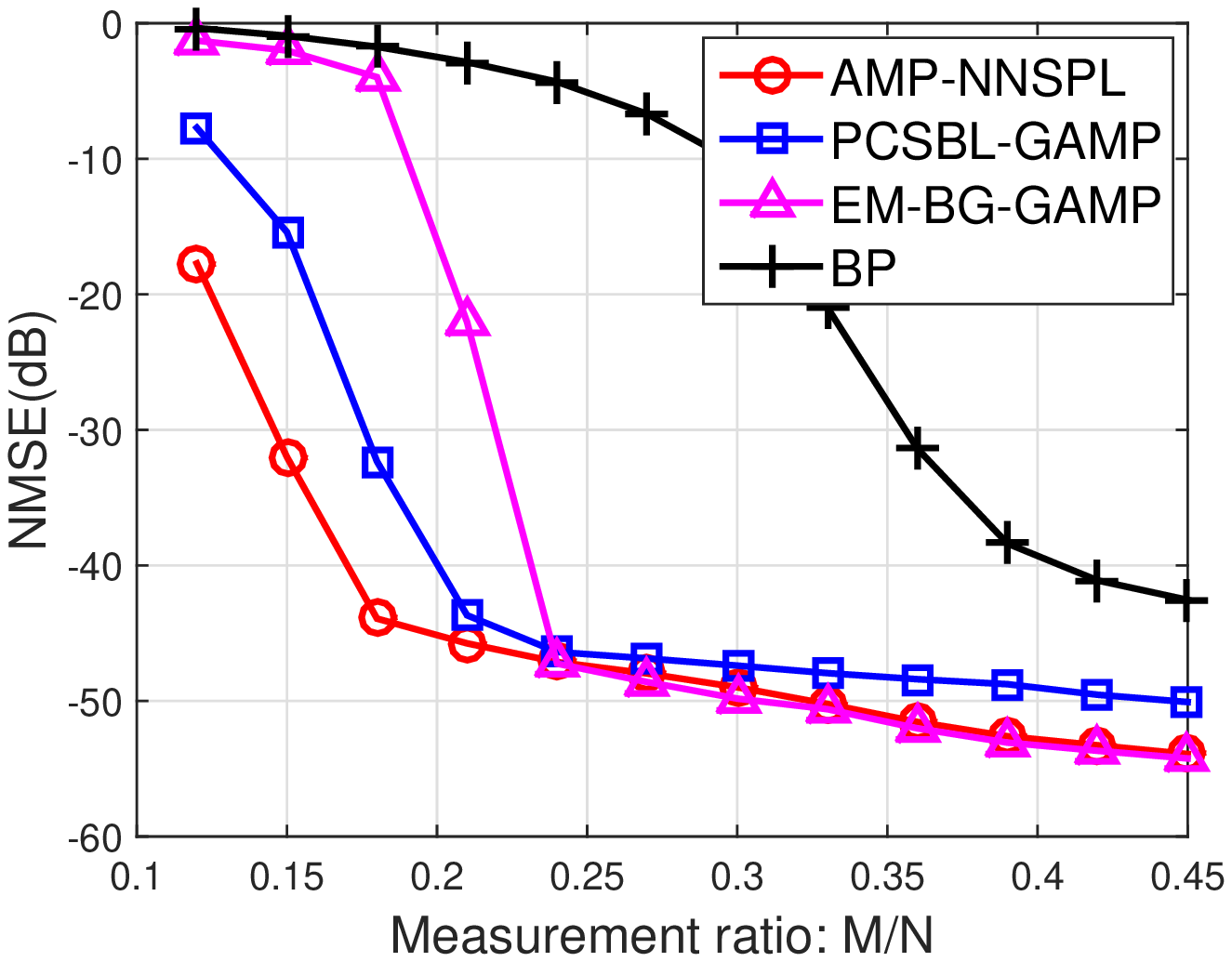}

\protect\caption{Success rate (left) in noiseless case and NMSE (right) at $SNR=50\textrm{dB}$
vs. $M/N$ for real 2D angiogram image. }
\label{2D_image}
\end{figure}

\begin{figure}[H]
\begin{centering}
\includegraphics[width=7cm]{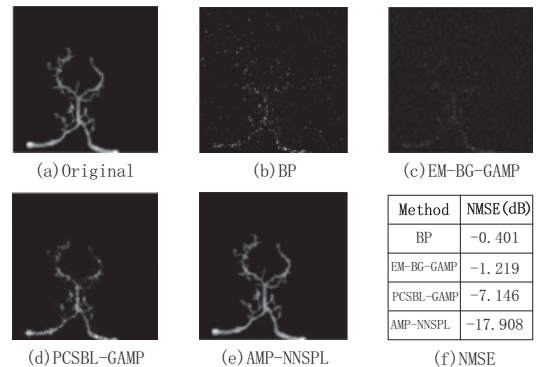}
\par\end{centering}

\protect\caption{Recovery results of real 2D angiogram image in noisy setting when
$M/N=0.12$ and $SNR=50$ dB.}

\label{2D_image-1}
\end{figure}

To evaluate the performance on real data, we consider a real angiogram
image\cite{hegde2015nearly} of 100$\times$100 pixels with sparsity
around 0.12. Fig. \ref{2D_image} depicts the success rate in noiseless
case and NMSE at $SNR=50$ dB, respectively. The MBCS-LBP and PC-SBL
algorithms are not included due to their high computational complexity.
It can be seen that AMP-NNSPL significantly outperforms other methods
both in terms of success rate and NMSE. In particular, when $M/N=0.12$
and $SNR=50$ dB, typical recovery results are illustrated in Fig.
\ref{2D_image-1}, which shows that AMP-NNSPL achives the best reconstruction
performance.

\section{\label{sec:Conclusion}Conclusion}

In this lettter, we propose an efficient algorithm termed AMP-NNSPL
to recover clustered sparse signals when the sparsity pattern is unknown.
Inspired by the k-NN algorithm, AMP-NNSPL learns the sparse ratios
in each AMP iteration as the average of their nearest neighbor estimates
using EM, thereby the sparsity pattern is learned adaptively. Experimental
results on both synthetic and real data demonstrate the state-of-the-art
performance of AMP-NNSPL.

\newpage{}

\bibliographystyle{IEEEtran}
\bibliography{IEEEabrv,Mybib}

\end{document}